\documentclass[preprint,epsfig,onecolumn,floats,showpacs]{revtex4}
\usepackage{amssymb}
\usepackage{graphicx}
\usepackage{epsfig}
\usepackage{amsmath}
\usepackage{amsfonts}

\begin{document}

\title{Quantum computation with graphene nanoribbon}
\author{Guo-Ping Guo}
\email{gpguo@ustc.edu.cn}
\author{Zhi-Rong Lin}
\author{Xiao-Peng Li}
\author{Tao Tu}
\email{tutao@ustc.edu.cn}
\author{Guang-Can Guo}
\affiliation{Key Laboratory of Quantum Information, University of Science and Technology
of China, Chinese Academy of Sciences, Hefei 230026, People's Republic of
China}

\begin{abstract}
We propose a scalable scheme to implement quantum computation in graphene
nanoribbon. It is shown that electron or hole can be naturally localized in
each zigzag region for a graphene nanoribbon with a sequence of Z-shaped
structure without exploiting any confined gate. An one-dimensional graphene
quantum dots chain is formed in such graphene nanoribbon, where electron or
hole spin can be encoded as qubits. The coupling interaction between
neighboring graphene quantum dots is found to be always-on Heisenberg type.
Applying the bang-bang control strategy and decoherence free subspaces
encoding method, universal quantum computation is argued to be realizable
with the present techniques.
\end{abstract}

\date{\today}
\maketitle

%-------------------------------------------------------------------------------

Electron spin is one of the leading candidates for the realization of a
practical solid qubit~\cite{Loss1998}. The coherent manipulation of electron
spins in GaAs quantum dots has been efficiently realized~\cite%
{Petta,Koppens2006}. However due to the interaction with the environment,
the decoherence time is often in nanoseconds scale in GaAs quantum dots~\cite%
{Hanson,Petta,Koppens2005}. Even by applying the complex technique to
prepare nuclear state, the dephasing time for spin qubits is just about 1$%
\mu s$~\cite{Reilly}. The decoherence is one of the most challenges in the
way to quantum computer in GaAs quantum dots. Due to the weak spin-orbit
coupling and hyperfine interactions in carbon, graphene is argued to be an
excellent candidate for quantum computation~\cite{Loss2007}. However, due to
the special band structure of graphene~\cite{Neto}, the low-energy
quasiparticles in graphene behave as Dirac fermions, and the Klein tunneling
and Chiral effect lead to the fact that it is non-trivial to form good
quantum dot (localized electron states) in graphene. It has been shown that
the massless Dirac fermions in graphene can ben confined by using suitable
transverse states in graphene nanoribbons (GNR)~\cite{Loss2007,Silvestrov},
by combining single and bilayer regions of graphene \cite{Nilsson,Peeters}
or by using inhomogeneous magnetic fields~\cite{Martino}. Recently, there
was an experiment report that GNR with well defined zigzag or armchair edge
structures can be chemically produced~\cite{HJDai}. It has also been
discovered that localized states exist in the zigzag region in Z-shaped GNR~%
\cite{Wang}.

Here we present a scalable quantum computation scheme based on Z-shaped GNR
quantum dot system without exploiting any confined gates. The localized
particle can be chosen to be electron or hole by adjusting the back gate
even in the room temperature. The qubit is encoded on the electron (hole)
spin states localized in the zigzag region of the GNR with a sequence of
Z-shaped structure. The interaction between qubits is determined by the GNR
geometrical structure and found to be Heisenberg form. By exploiting
bang-bang (BB) control strategy and decoherence free subspaces (DFSs)
encoding method, universal quantum gates are shown to be realizable in this
system with the present techniques.

Based on the $\pi $ orbital tight-binding model, the local density of state
(LDOS) and the band structure of the zigzag region in a GNR with a sequence
of Z-shaped structure can be obtained by the direct diagonalization of the
single particle Hamiltonian $H_{0}=\sum_{ij}{\tau _{ij}}|i\rangle \langle j|$%
, where the hopping matrix element $\tau _{ij}=-\tau $ if the orbits $%
|i\rangle $ and $|j\rangle $ are nearest neighboring on the honeycomb
lattice, otherwise $\tau _{ij}=0$ \cite{Brey,Nakada}. From the calculated
band structure, we can see that there are several localized states with
electron-hole symmetry around the zero energy point as shown in Fig.~\ref%
{figure1}a. Thus we can choose to get one localized electron or hole in the
zigzag region by adjusting the Fermi level through the back gate. The
electron ground state energy and the energy gap between the ground state and
the first excitation state are very sensitive to the size of the zigzag
region, as shown in Fig~\ref{figure2}. It has been known that the width of
the armchair GNR ($N$ unit cells) decides whether the system is metallic or
semiconducting~\cite{Brey,Nakada}. If $N=3m-1$ ($m$ being an integer), the
armchair GNR is metallic, otherwise it is semiconducting. In addition, for
the present Z-shaped structure the boundaries along the ribbon of armchair
region is unsymmetrical when $N$ is even. Actually, in our calculation we
find there is no confined state in the zigzag region of Z-shaped GNR when $%
N=3m-1$ or $N=2m$ as shown in Fig.~\ref{figure2}. On the other hand, when $N
$ is $7$ and the length of the zigzag region $L$ (unit cells) (see Fig. \ref%
{figure1}c) is $3,4,5,6,7$ and $N=9$, $L=3,4$ both the ground level and
energy gap are above $0.1$eV. Thus we can confine electron (hole) to form
quantum dot even in the room temperature.

\begin{figure}[tbp]
\centering
\includegraphics[width=5in]{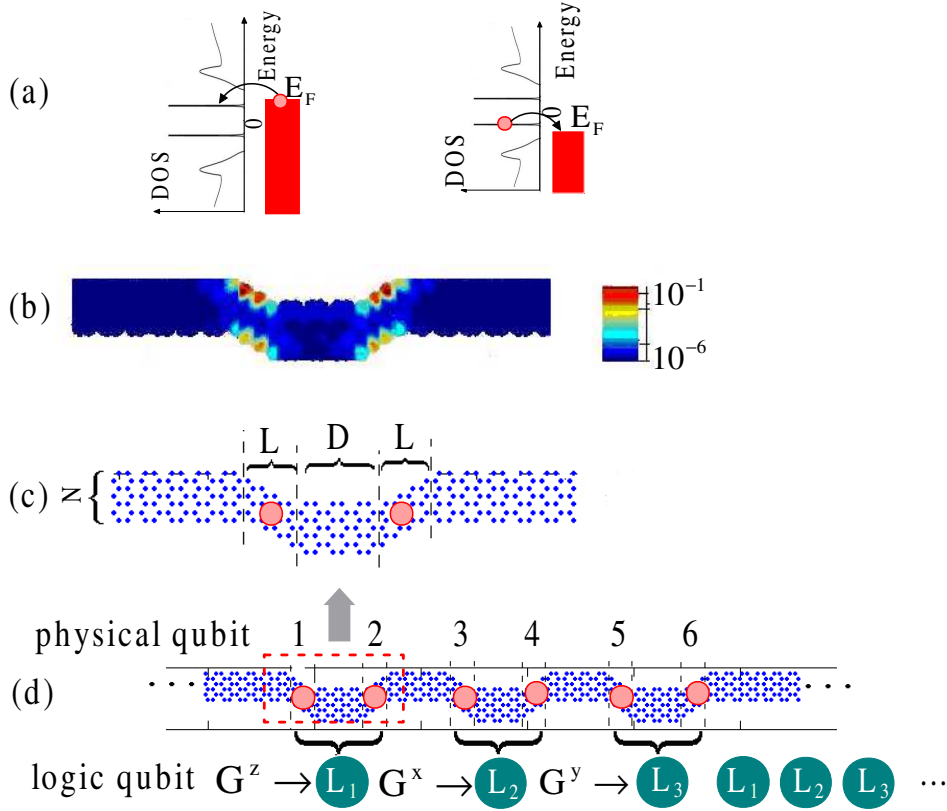}
\caption{(Color online) Schematic of the proposed architecture of GNR for
quantum computation. (a) The Z-shaped GNR quantum dot can localize one
electron (left figure) or hole (right figure) in the zigzag region by
adjusting the Fermi level through the back gate. (b) Local density of states
of GNR with two Z-shaped structure in series. (c) A GNR with two Z-shaped
structure in series, each zigzag region confines one electron. The physical
qubit is encoded into the spin of the confined electron. (d) Special
encoding method to eliminate the interaction between logical qubits.
Physical qubits $1$ and $2$ form logical qubit $L_{1}$; physical qubits $3$
and $4$ form logical qubit $L_{2}$; physical qubits $5$ and $6$ form logical
qubit $L_{3}$. The $G^{z}$ , $G^{x}$, $G^{y}$ are the BB operation sets of $%
L_{1}$, $L_{2}$ and $L_{3}$ respectively. }
\label{figure1}
\end{figure}

\begin{figure}[tbp]
\centering
\includegraphics[width=4.0in]{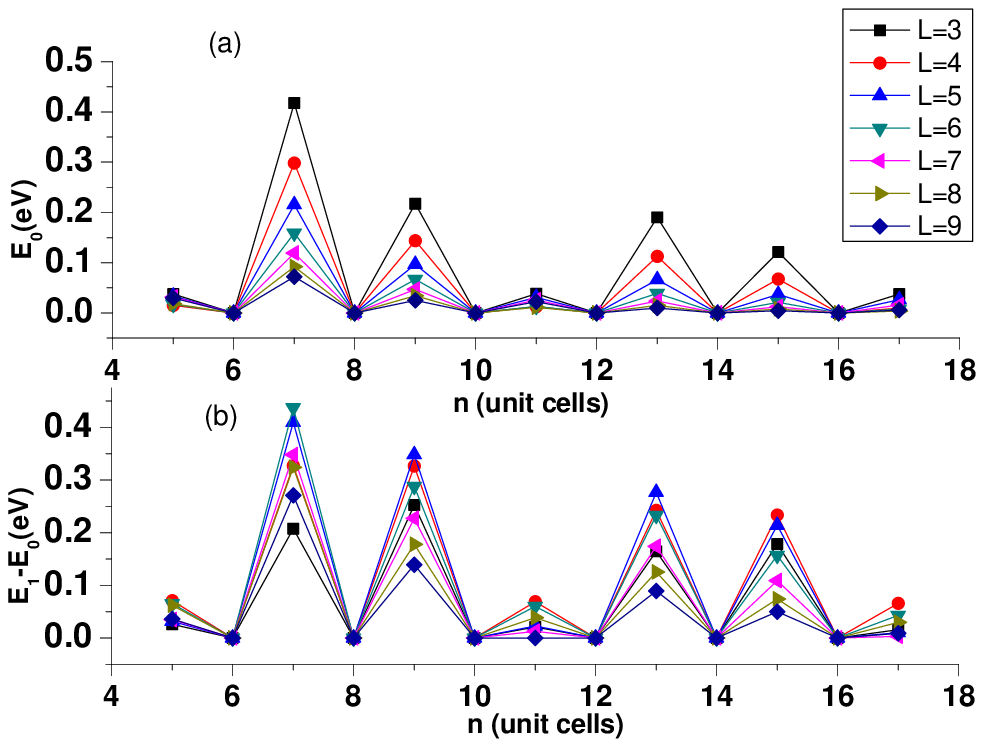}
\caption{(Color online) The ground energy level ($E_{0}$) and energy gap ($%
E_{1}-E_{0}$) between the ground state and the first excitation state of the
Z-shaped GNR quantum dot device with different width of nanoribbon ($N$ unit
cells) and different length of quantum dot region ($L$ unit cells).}
\label{figure2}
\end{figure}

\begin{figure}[tbp]
\centering
\includegraphics[width=4in]{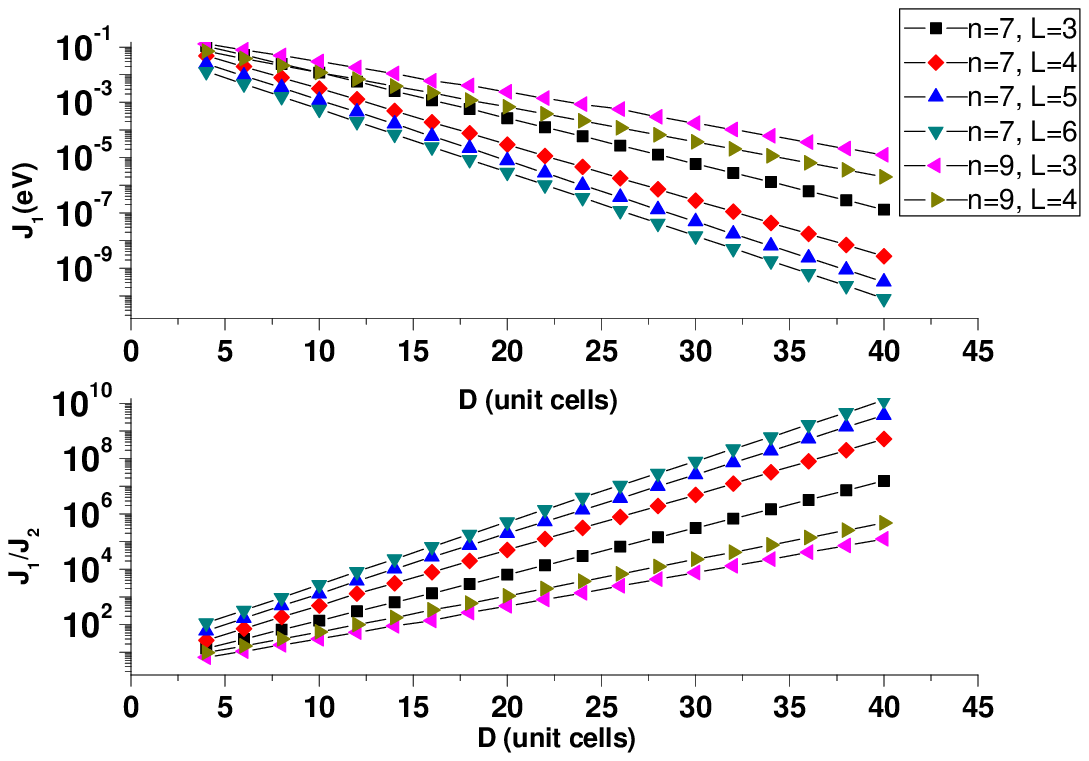}
\caption{(Color online) (a) The coupling energy $J_{1}$ of two nearest
neighboring qubits dependence on the number of unit cells between two qubits
is presented for different size of quantum dot region. (b) The ratio of $%
J_{1}$ to the next nearest neighboring exchange energy $J_{2}$ depend on the
distance $D$ of two neighboring qubits for different size of quantum dot
region.}
\label{figure3}
\end{figure}

Fig.~\ref{figure1}b shows the spatial distribution of local density of
ground state for a GNR with two Z-shaped structure in series. Each zigzag
region (quantum dot) confines one electron and the quantum dots are coupled
by the exchange coupling $J_{1}$. We can obtain $J_{1}$ by calculating the
exchange integral $J_{1}=\int \varphi _{1}^{\ast }(\vec{r_{1}})\varphi
_{2}^{\ast }(\vec{r_{2}})\frac{e^{2}}{|\vec{r_{1}}-\vec{r_{2}}|}\varphi _{1}(%
\vec{r_{2}})\varphi _{2}(\vec{r_{1}})d\vec{r_{1}}d\vec{r_{2}}$, where $%
\varphi _{1}(\vec{r})$ and $\varphi _{2}(\vec{r})$ are the wavefunction of
neighboring graphene quantum dots. We can also calculate the next nearest
neighboring exchange coupling $J_{2}$ by the same method. Obviously, the
exchange coupling $J_{1}$, $J_{2}$ are determined by the geometrical
structure of the nanoribbon. For each $N$ and $L$, $J_{1}$ and $J_{1}/J_{2}$
depend on the number of unit cells ($D$) between two neighboring qubits. By
numerical calculations, $J_{1}$, $J_{1}/J_{2}$ are obtained with different $%
N $, $L$, $D$, as shown in Fig.~\ref{figure3}. For $N=7$, $L=6$, $D=18$, $%
J_{1}=8\mu $eV, $J_{1}/J_{2}=10^{5}$, we can safely neglect this non-nearest
neighboring qubits coupling. For clarity, in the following discussion we
focus on the atomic structure with $N=7$, $L=6$, $D=18$.

To carry out quantum computation, we have to form the logical qubit and
realize universal quantum gate. It has been shown that single qubit
rotations combined with two-qubit operations can be used to create basic
quantum gates~\cite{DiVincenzo}. The spin of the localized electron or hole
can be used as the physical qubit and the GNR with a sequence of Z-shaped
structure forms an one-dimensional qubit chain as shown in Fig.~\ref{figure1}%
d. The neighboring qubits in this chain have an always-on Heisenberg
interaction $H=J_{1}\vec{S_{1}}\cdot \vec{S_{2}}$. Here $\vec{S_{1}}$ and $%
\vec{S_{2}}$ are the spin operator of the neighboring localized electron
(hole). It has been known that BB control strategy and DFSs encoding method
do not require directly controlling the interaction between qubits~\cite%
{Benjamin,Zhang}. The quantum information in qubits can be protected from
decoherence induced by the environment and undesired disturbance induced by
the inherent qubit-qubit interaction with these strategies.

For a sequence of Z-shaped structure GNR with $N=7$, $L=6$, $D=18$, the
Hamiltonian of the system can be expressed as
\begin{equation}
H_{I}=\sum_{i,j}J_{i,j}(\sigma _{i}^{x}\otimes \sigma _{j}^{x}+\sigma
_{i}^{y}\otimes \sigma _{j}^{y}+\sigma _{i}^{z}\otimes \sigma _{j}^{z}),
\label{Hamiltonian}
\end{equation}%
where $\sigma _{i,j}^{x,y,z}$ are the spin Pauli operators of the localized
electron (hole) in the quantum dots, $i$ and $j$ represent two neighboring
dots. Here we have neglected the interaction between non-neighboring dots,
which has been shown to be $5$ orders smaller than the neighboring
interaction.

To avoid the spin qubits to entangle with the environment, we can apply a BB
operation $U_{z}=exp(-i\sigma _{z}\pi /2)$ to each quantum dot region. Such
rotation operations can be realized if a pulsed magnetic field could be
applied exclusively~\cite{Loss1998}. To counteract phase decoherence, we can
use DFSs encoding~\cite{Duan}. For a simply DFSs encoding, two physical
qubits can encode a logical qubit:
\begin{equation}
|0\rangle _{L}=|\uparrow _{1}\downarrow _{2}\rangle ,|1\rangle
_{L}=|\downarrow _{1}\uparrow _{2}\rangle .  \label{Enstate1}
\end{equation}%
As shown in Fig.~\ref{figure1}c, we use localized electron in the two
neighboring zigzag regions to form a logical qubit.

In order to protect quantum information in the logical qubits, we must
decouple the always-on Heisenberg interaction between two physical qubits
within a logical qubits and interaction between two neighboring logical
qubits. A nonsynchronous BB pulse operations and a special encoding method
can be exploited to eliminate these interactions~\cite{Zhang}. Here we
propose an architecture in which the one-dimensional GNR chain form a
periodic structure $L_{1}L_{2}L_{3}L_{1}L_{2}L_{3}$ $\cdots $ with three
logical qubits as a unit, as shown in Fig.~\ref{figure1}d. $L_{1}$
represents a logical qubit encoded as Eq.(\ref{Enstate1}). $L_{2}$ is a
logical qubit encoded as
\begin{equation}
|0\rangle _{L2}=\frac{1}{2}(|\uparrow \rangle _{3}+|\downarrow \rangle
_{3})(|\uparrow \rangle _{4}-|\downarrow \rangle _{4}),
\end{equation}%
\begin{equation}
|1\rangle _{L2}=\frac{1}{2}(|\uparrow \rangle _{3}-|\downarrow \rangle
_{3})(|\uparrow \rangle _{4}+|\downarrow \rangle _{4}).
\end{equation}%
And $L_{3}$ is a logical qubit encoded as
\begin{equation}
|0\rangle _{L3}=\frac{1}{2}(|\uparrow \rangle _{5}+i|\downarrow \rangle
_{5})(|\uparrow \rangle _{6}-i|\downarrow \rangle _{6}),
\end{equation}%
\begin{equation}
|1\rangle _{L3}=\frac{1}{2}(|\uparrow \rangle _{5}-i|\downarrow \rangle
_{5})(|\uparrow \rangle _{6}+i|\downarrow \rangle _{6}).
\end{equation}%
With this periodic architecture, we have to apply nonsynchronous BB pluse
operations respectively to $L_{1}$, $L_{2}$, $L_{3}$ from the operation set $%
G^{z}=\{I,U_{z},R_{z}\}$, $G^{x}=\{I,U_{x},R_{x}\}$, $G^{y}=\{I,U_{y},R_{y}%
\} $, where $U_{z}=-\sigma _{1}^{z}\otimes \sigma _{2}^{z}$, $%
R_{z}=-iI_{1}^{z}\otimes \sigma _{2}^{z}$, $U_{x}=-\sigma _{1}^{x}\otimes
\sigma _{2}^{x}$, $R_{x}=-iI_{1}\otimes \sigma _{2}^{x}$, $U_{y}=-\sigma
_{1}^{y}\otimes \sigma _{2}^{y}$, and $R_{y}=-iI_{1}\otimes \sigma _{2}^{y}$%
. Then we obtain a quantum computation system with entirely decoupled
logical qubits.

\begin{figure}[tbp]
\centering
\includegraphics[width=4in]{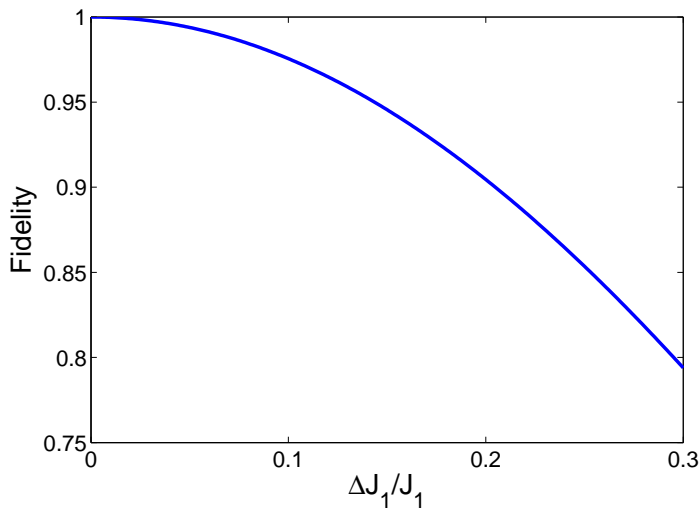}
\caption{The fidelity of $\bar{X}$ gate against the fluctuation or
inhomogeneity of the exchange coupling energy $J_{1}$. }
\label{figure4}
\end{figure}

Now we show how to carry out universal quantum gates of the logical qubits
defined above. Logical operations $\bar{X}$ and $\bar{Z}$ can generate all
SU(2) transformations of logical qubit. For logical qubit $L_{1}$, $\bar{X}=%
\frac{1}{2}(\sigma _{1}^{x}\otimes \sigma _{2}^{x}+\sigma _{1}^{y}\otimes
\sigma _{2}^{y})$, $\bar{Z}=\frac{1}{2}(\sigma _{1}^{z}-\sigma _{2}^{z})$. $%
\bar{X}$ can be easily achieved by recoupling qubits $1$ and $2$ by
adjusting the BB pulses of both qubits to be synchronous~\cite{Zhang}. The
operation time can be obtained by $J\Delta t=\hbar \pi /4$ , for $N=7$, $L=6$
, $D=18$, $\Delta t=0.2$ns. $\bar{Z}$ can be achieved by directly varying
the Zeeman splitting on the two physical qubits individually analogous to
single-qubit operations in the Loss-DiVincenzo quantum computer~\cite%
{Loss1998}. The operation time of this $\bar{Z}$ gate can be about $1$ns
when $20$mT magnetic field could be pulsed exclusively onto each quantum dot
region. The fidelity of the $\bar{X}$ gate can be effected by the
fluctuation or inhomogeneity in the exchange coupling $J_{1}$ between
different dots. The charge noise in the back gate can cause the fluctuation
of $J_{1}$. The main sources of the $J_{1}$ inhomogeneity between different
dots are the disorder, irregular edges and defect of the GNR. Short-range
disorder scarcely affects the LDOS of the ground state. The irregular edge
effect and long-range disorder change the LDOS but do not destroy the
confined states~\cite{Wang}. To get high fidelity for $\bar{X}$ operation,
we should avoid the long-range disorder and irregular edge. Actually, if we
know the coupling $J_{1}$ between different dots exactly, inhomogeneity of $%
J_{1}$ can not effect the fidelity of $\bar{X}$ gate when corresponding
inhomogeneous operation times are used. In addition, we find the effect of
the $J_{1}$ inhomogeneity or fluctuation to the fidelity of the $\bar{X}$
gate is small as shown in Fig.~\ref{figure4}. Because the nuclear field
would change the evolution of the spin states, the fidelity of the $\bar{Z}$
gate is dominated by the nuclear field~\cite{Koppens2006}. The fidelity of
the $\bar{Z}$ gate can be very high due to small nuclear field in graphene
system. Similarly, high fidelity operation $\bar{X}$ and $\bar{Z}$ can be
also realized for logical qubits $L_{2}$ and $L_{3}$.

We can construct CNOT gate between two neighboring logical qubits, for
example $L_{1}$ and $L_{2}$, by W gate $W=|0\rangle \langle 0|\otimes
I+|1\rangle \langle 1|\otimes e^{i2\theta \bar{Z}}=e^{i\theta \bar{Z}\otimes
\bar{Z}}$ and Hadamard operation~\cite{Bremner}. By performing Hadamard
transformation
\begin{equation}
H=\frac{1}{\sqrt{2}}\left[
\begin{array}{cc}
1 & 1 \\
1 & -1%
\end{array}%
\right] ,
\end{equation}%
to the two physical qubits of the second logical qubit $L_{2}$ and changing
the BB control pulse to be the same with $L_{1}$, we can recouple the two
neighboring logical qubits and implement W gate of logical qubits of $L_{1}$
and $L_{2}$. For the present graphene quantum dots chain with $N=7$, $L=6$, $%
D=18$, the total operation time of a CNOT gate can be implemented in about $%
1 $ns with an oscillating magnetic field of $100$mT to achieve the Hadamard
operation. Similar to the above discussion for $\bar{X}$ and $\bar{Z}$
operation, we can find that the fluctuation or inhomogeneity of $J_{1}$ and
the nuclear field have trivial effect to the fidelity of the CNOT gate in
the present protocol.

The major decoherence sources of spin qubits in solid state system have been
identified as the spin-orbit interaction and hyperfine interaction. The weak
spin-orbit coupling have been predicted in carbon material due to the low
atomic weight~\cite{Min}. Since the primary component of natural carbon is
the zero spin isotope $^{12}C$, the very long coherence time given by
hyperfine coupling has been theoretically argued~\cite{Loss2007}. Assuming
the abundance of $^{13}C$ is about $1\%$ as in the nature carbon material,
the decoherence time has been predicted to be more than $10\mu$s in the
graphene quantum dot~\cite{Loss2007,Coish}. This decoherence time is $4$
orders longer than the gates operation time of the present protocol. In
addition, the decoherence time can be much longer if the percentage of $%
^{13}C$ is decreased by isotopic purification.

In this paper we have presented a scalable scheme of quantum computation
based on GNR with a sequence of Z-shaped structure. No confined gates is
needed to localize the particle, which can be chosen to be electron or hole
by adjusting back gate. The qubit is encoded in electron or hole spin
states, which is naturally localized in the zigzag region of GNR even in
room temperature. The neighboring qubits are found to have an always-on
Heisenberg interaction and the dynamical decoupling techniques with DFSs is
exploited to achieve universal quantum computation in this system. Due to
recent achievement in production of graphene nanoribbon, this proposal may
be implementable within the present techniques.

We thank Prof. Q. W. Shi and Dr. Z. F. Wang for helpful discussions. This
work was funded by National Fundamental Research Program, the Innovation
funds from Chinese Academy of Sciences, NCET-04-0587, and National Natural
Science Foundation of China (Grant No. 60121503, 10574126, 10604052).

\end{document}